\documentclass[sigconf]{acmart}

\usepackage{booktabs} 

\setcopyright{none}

\acmDOI{}

\acmISBN{}

\acmConference{}{}{}
\acmYear{}
\copyrightyear{}

\acmArticle{}
\acmPrice{}

\begin{document}
\title{Mitigating Cold Starts in Serverless Platforms}
\subtitle{A Pool-Based Approach}

\author{Ping-Min Lin}
\affiliation{\institution{Carnegie Mellon University}}
\email{pingminl@cs.cmu.edu}

\author{Alex Glikson}
\affiliation{\institution{Carnegie Mellon University}}
\email{aglikson@cs.cmu.edu}

\begin{abstract}
Rapid adoption of the 'serverless' (or Function-as-a-Service, FaaS) paradigm \cite{childers18}, pioneered by Amazon with AWS Lambda and followed by numerous commercial offerings and open source projects, introduces new challenges in designing the cloud infrastructure, balancing between performance and cost. While instant per-request elasticity that FaaS platforms typically offer application developers makes it possible to achieve high performance of bursty workloads without over-provisioning, such elasticity often involves extra latency associated with on-demand provisioning of individual runtime containers that serve the functions. This phenomenon is often called 'cold starts' \cite{tresness18}, as opposed to the situation when a function is served by a pre-provisioned 'warm' container, ready to serve requests with close to zero overhead. Providers are constantly working on techniques aimed at reducing cold starts. A common approach to reduce cold starts is to maintain a pool of 'warm' containers, in anticipation of future requests. In this project, we address the cold start problem in serverless architectures, specifically under the Knative Serving FaaS platform. We implemented a pool of function instances and evaluated the latency compared with the original implementation, which resulted in an 85\% reduction of P99 response time for a single instance pool.

\end{abstract}

\maketitle

\section{Introduction}

\subsection{Serverless Platforms}
Serverless (also known as Function-as-a-Service, FaaS) is an emerging paradigm of cloud computing that allows developers to abstract away underlying infrastructure down to the "function" level of an application. Compared to traditional cloud offerings such as Platform-as-a-Service (PaaS), serverless allows the developer to focus on the functionality of the service itself without worrying about any environment and system issues, such as scaling and fault tolerance. Most cloud service providers now support some degree of the serverless architecture, e.g. AWS Lambda \cite{awslambda}, Google Cloud Functions \cite{gcf} etc., and various open source frameworks are released, including IBM incubated OpenWhisk \cite{openwhisk} and Knative \cite{knative}. The back-end of these services vary from existing generic container technology \cite{openwhisk, knative} to highly customized virtualization methods\cite{firecracker}, but all aim to provide a simple-to-deploy model for developers.

\subsection{Knative and Knative Serving}
Knative\cite{knative} is an open source initiative aiming to provide a platform for developing container applications on top of the Kubernetes container orchestration platform. It offers ready-to-use components built with Kubernetes Custom Resource Definitions (CRDs) for developers to build, deploy and scale their functions end-to-end from source code. The project currently provides three components:
\begin{enumerate}
    \item \textbf{Build} {an end-to-end source-to-URL deployment tool for container images.}
    \item \textbf{Eventing} {management and delivery of events between containers.}
    \item \textbf{Serving} {a request-driven autoscaling module that can scale to zero.}
\end{enumerate}

Knative also leverages the service mesh networking framework Istio for efficient and flexible traffic routing. These components enable Knative to satisfy the requirements of a serverless framework: scaling dynamically (including scaling to zero) and easy to deploy.

\begin{figure}[t]
    \centering
    \includegraphics[width=0.4\textwidth]{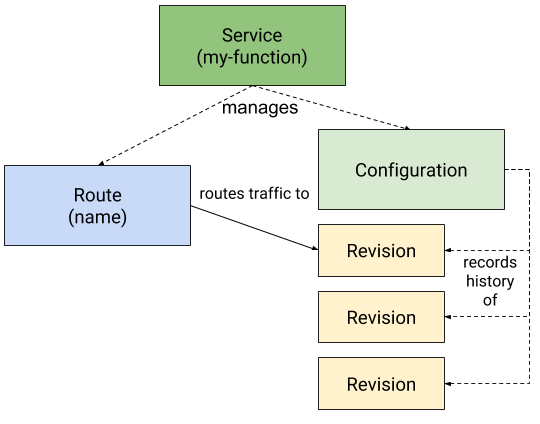}
    \caption{Knative Control Flow, image obtained from the Knative/Serving repository \cite{knativeserving}}
    \label{fig:knativectrlflow}
\end{figure}

Knative Serving \cite{knativeserving} takes care of scaling and load-balancing of the serverless functions in Knative. A high level hierarchy of the components in the Serving project is shown in Figure \ref{fig:knativectrlflow}. Functions are represented by container images, where each function instance runs as a pod/container. Knative Serving provides high level abstractions on top of Kubernetes to allow service administrators to easily roll out new services. A Knative \textit{service} is a management object for functions which are defined by configurations. Any updates in the \textit{configuration} will create a new immutable \textit{revision} representing the state of the function. Each \textit{service} also points to a specific \textit{route} that keeps information of how the requests should be redirected.

\subsection{Cold Start Problem}
Ideally, we would like to have minimal overhead when invoking the functions.  However, when the platform needs to spin up the first instance, the underlying resources still needs time for initialization. This bootstrapping also happens when the autoscaler provisions additional instances to handle traffic. The initialization time is unavoidable for each function instance and introduces a delay until the instance can respond to request(s). This issue is very common in  serverless platforms, and is known as the \textbf{cold start problem}. Although FaaS offering typically suffer from cold-starts, the overhead each platform incurs varies based on the underlying implementation of the functions.

Knative is built on top of Kubernetes and uses containers as the computational resource. The cold start overhead could be split into two categories:
\begin{enumerate}
    \item Platform overhead: Overheads introduced due to Knative itself, this includes network bootstrapping, network sidecar injection, pod provisioning, etc. This delay is rather uniform throughout different functions.
    \item Application-dependent initialization overhead: Any overhead caused by the application, e.g. model initialization for a machine learning application. This type of latency varies between different applications.
\end{enumerate}
For example, in our experiments with Knative, cold starts for a simple HTTP server caused the application to wait around 5 seconds before the first response, while the delay introduced to a naive image classifier implemented in Tensorflow can be up to 40 seconds. This latency is rather intolerable, especially when the response time for a typical web service is in the order of milliseconds.

Our approach to solve the cold start problem is to maintain a pool of warm pods that will stand-by and be immediately available to functions with increasing demand, eliminating the cold start overhead by provisioning the pods beforehand. This pool can also be shared among multiple different services that use the same function.
In this report, we first describe the implementation of a pool system based on Knative Serving. We then evaluate the performance improvement of our solution compared to alternatives (e.g., Knative without a pool of pre-warmed Pods), and discuss results and insights. We also examine existing research of the cold start problem, and highlight some directions for future exploration.

\section{Implementation}
\subsection{Pool of Pods}
\begin{figure}[t]
    \centering
    \includegraphics[width=0.2\textwidth]{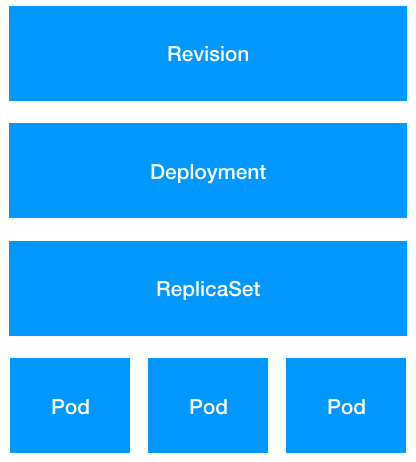}
    \caption{Knative Revision Component Structure}
    \label{fig:rev}
\end{figure}
\begin{figure}[ht]
    \centering
    \includegraphics[width=0.4\textwidth]{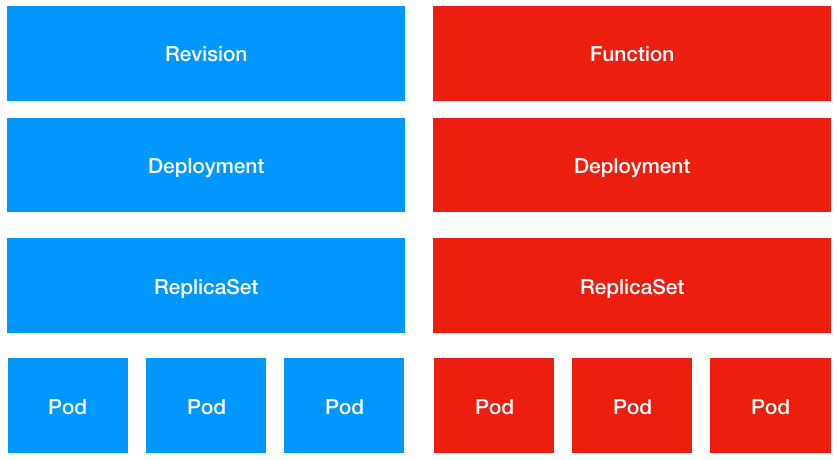}
    \caption{Our Pool Structure}
    \label{fig:revpool}
\end{figure}
\begin{figure}[ht]
    \centering
    \includegraphics[width=0.4\textwidth]{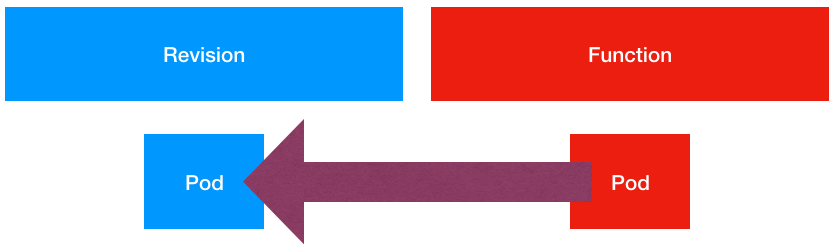}
    \caption{Migration of pods in the pool}
    \label{fig:migrate}
\end{figure}

We implement the pool as a Knative resource. In figure \ref{fig:rev}, we can see the underlying native components a \textit{revision} uses. We reused the \textit{revision} structure when defining the \textit{pool} resource, which can be seen in figure \ref{fig:revpool}. Using the same structure allows us to simplify the controlling methods and reduce the overhead of how labels propagate through the components. Since we followed the existing configurations in Knative using CRDs, the augmentation of the resource itself was pretty straightforward once we figure out the dependencies between the control logic. After defining the pool construct, we implement the logic of pod migration from the pool to the target revision (figure \ref{fig:migrate}).

\subsection{Pod Migration}
\begin{figure}[ht]
    \centering
    \includegraphics[width=0.45\textwidth]{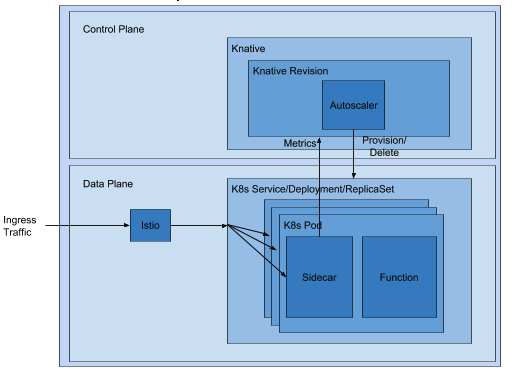}
    \caption{Original Autoscaling Logic}
    \label{fig:before}
\end{figure}

\begin{figure}[ht]
    \centering
    \includegraphics[width=0.45\textwidth]{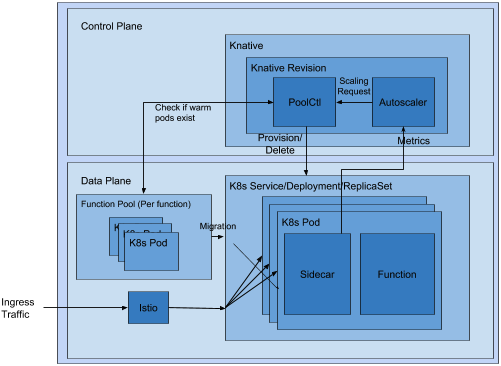}
    \caption{Our Autoscaling (with pool) Logic}
    \label{fig:after}
\end{figure}

The original logic of Knative's autoscaling is simple (figure \ref{fig:before}. Sidecar containers in the pod will collect and send traffic metrics to the autoscaler. The autoscaler will then scale up/down the revision directly by changing the number of desired pods in the underlying \textit{deployment}. This will trigger the reconciliation in Kubernetes to create/destroy pods in the \textit{replicaSet}. Our modification of the logic of the autoscaler can be seen in figure \ref{fig:after}: before changing the desired number of pods in the \textit{replicaSet}, we first check if there are any pods in the pool that can be migrated to the target. By a series of operations on the labels of these pods and selector of the \textit{replicaSets}, the pods are migrated to the target \textit{replicaSet}, and the stat collection for these pods are activated so the autoscaler can obtain correct metrics. If the number of new pods needed are more than the number of pods in the pool, the original logic kicks in and spawns an amount of new pods equal to the deficit between the desired number of new pods and the number of pods in the pool. This chain of operations is completed in a sequential order to reduce side effects causing race conditions. It currently takes around 2 seconds for the pods to get migrated.

The \textit{pool} resource definition took \textasciitilde200 LOC, and the migration code is \textasciitilde350 LOC to implement.

\section{Evaluation}
\subsection{Response Time to First Request}
We first evaluate our implementation by benchmarking the system with and without the pool of pods. Our system was built on a Google Cloud Platform n1-standard-4 (4 vCPUs, 15 GB memory) instance. We use two different applications for evaluation. The first application is a simple HTTP server implemented in Golang, and does some trivial computation before responding to the client. The second application is a image classifier for dogs and cats implemented with Python and Tensorflow, which needs to initialize and load the model (roughly 200MB in size) into memory before classification could be done. The results are shown in Table \ref{tab:rt}. We can see that the HTTP server saves approximately 7 seconds, which is over half of the cold start response time, and the delay of the image classifier has been cut down from almost 40 seconds to 7.5 seconds. This improvement is significant and we can see that the longer the application overhead is, the larger the difference pools make.
\begin{table}[ht]
  \caption{Response time (sec) (n=10)}
  \label{tab:rt}
  \begin{tabular}{ccc}
    \toprule
    Application&Cold Start($\sigma$)&Warm Start($\sigma$)\\
    \midrule
    HTTP Server & 12.123 (1.638)& 5.076 (1.055)\\
    Image Classifier & 39.25 (1.475)& 7.458 (0.641)\\
  \bottomrule
\end{tabular}
\end{table}
\subsection{Simulation on Traces}
\begin{figure}[ht]
    \centering
    \includegraphics[width=0.45\textwidth]{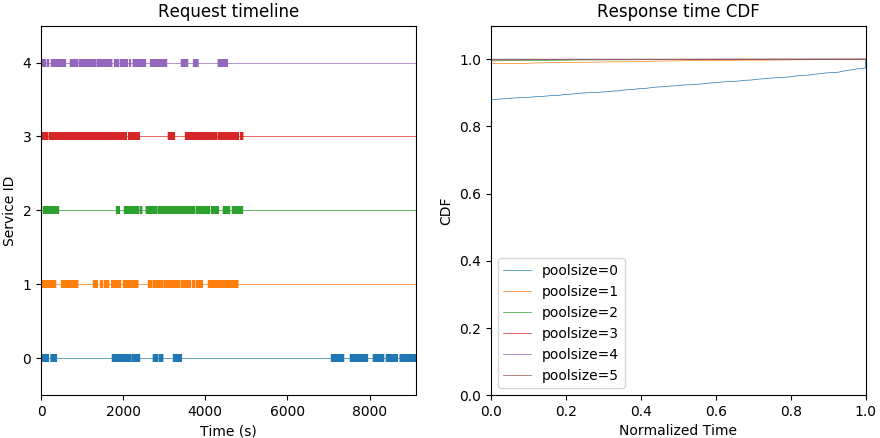}
    \caption{Sample trace and CDF of short application}
    \label{fig:7_30_sample}
\end{figure}

\begin{figure}[ht]
    \centering
    \includegraphics[width=0.45\textwidth]{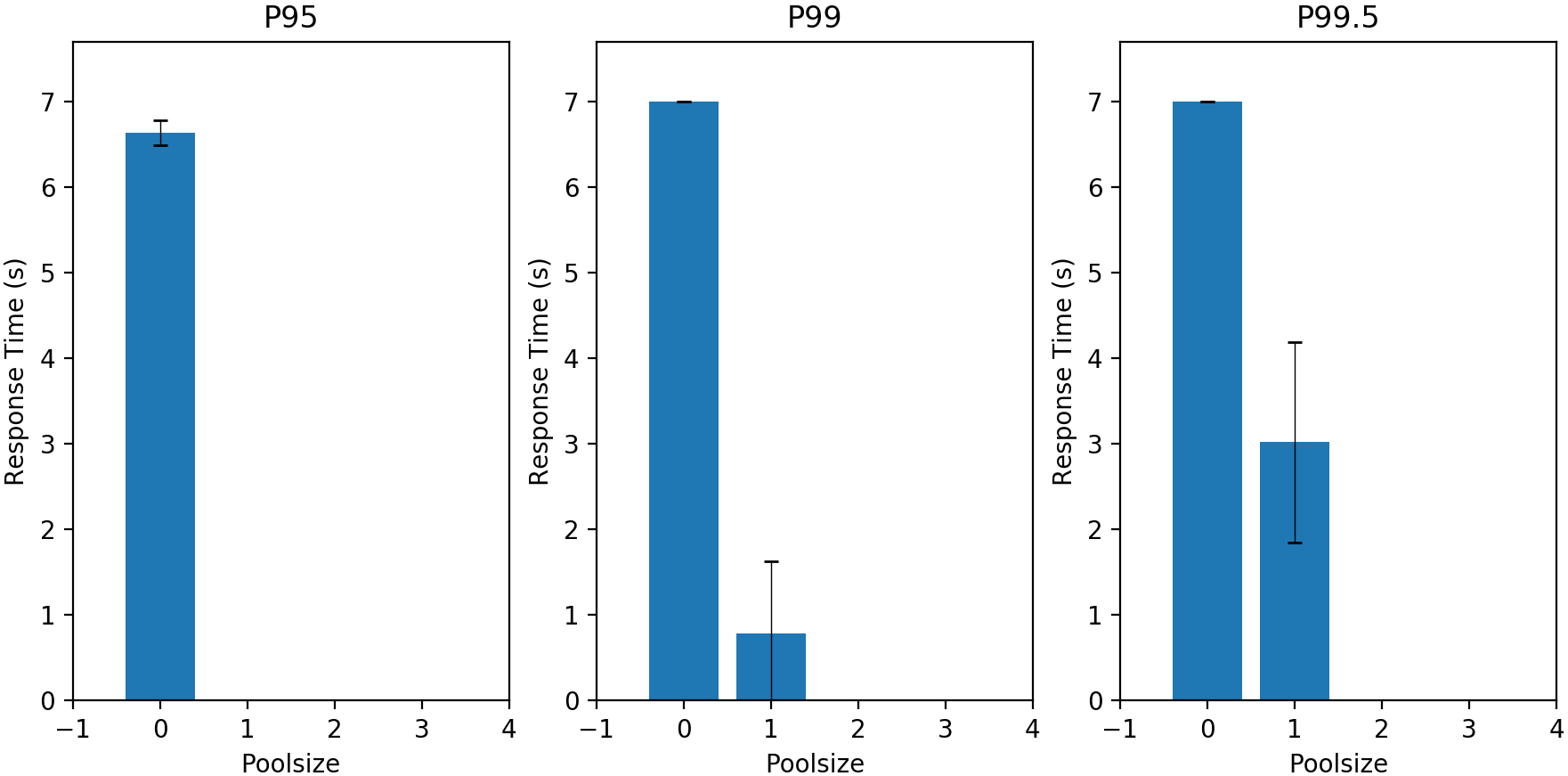}
    \caption{Percentiles of short application simulation}
    \label{fig:7_30_percentiles}
\end{figure}

\begin{figure}[ht]
    \centering
    \includegraphics[width=0.45\textwidth]{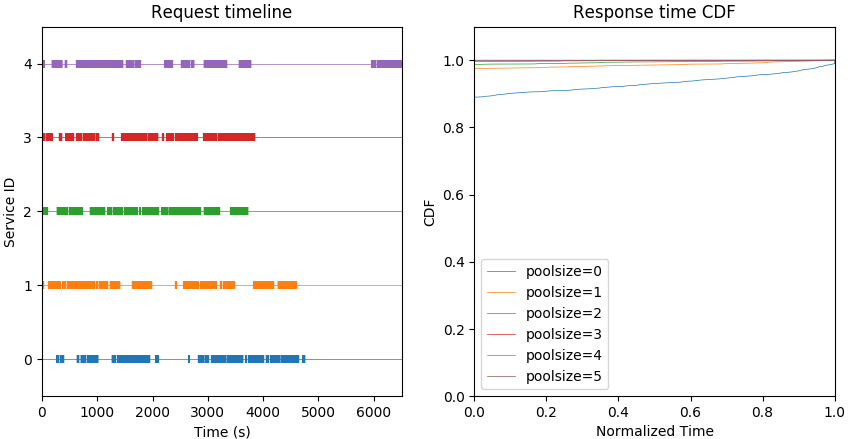}
    \caption{Sample trace and CDF of long application}
    \label{fig:32_60_sample}
\end{figure}

\begin{figure}[ht]
    \centering
    \includegraphics[width=0.45\textwidth]{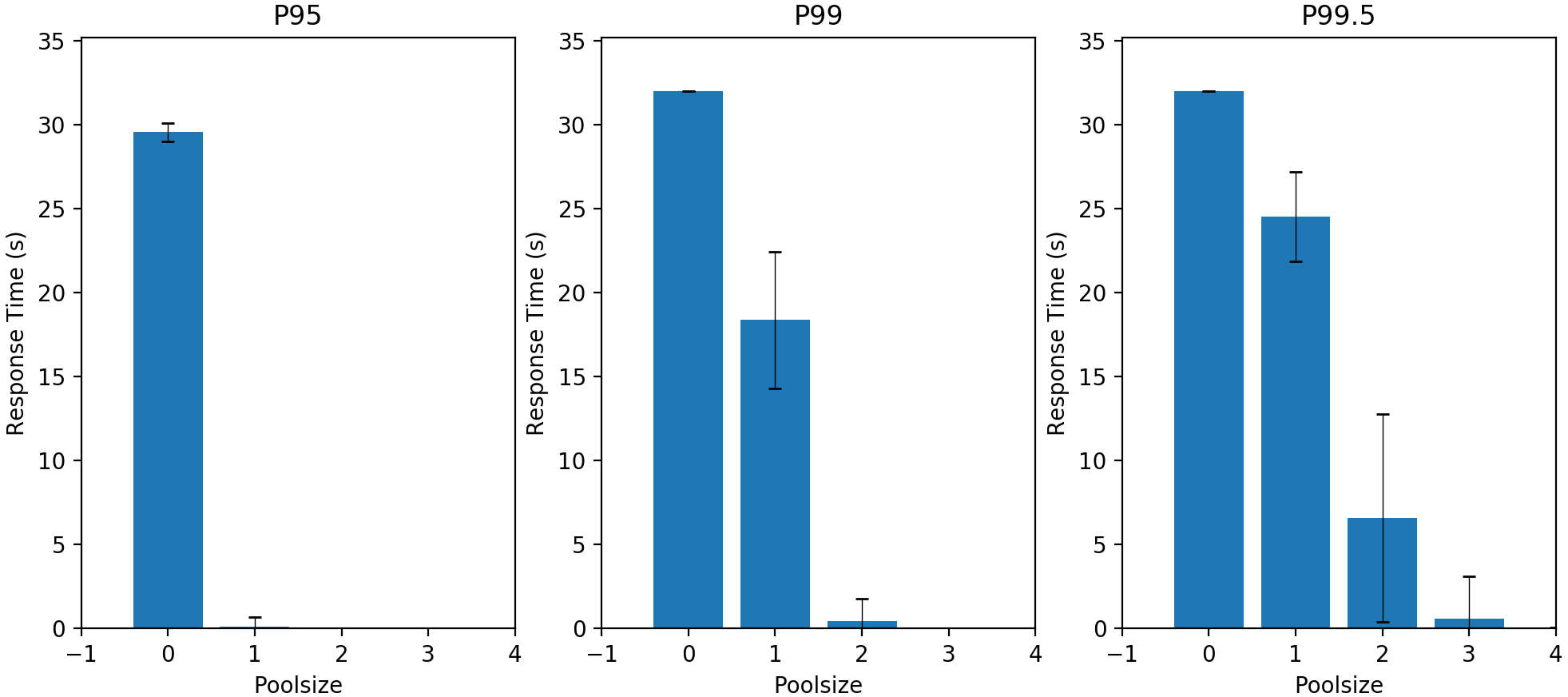}
    \caption{Percentiles of long application simulation}
    \label{fig:32_60_percentiles}
\end{figure}

In addition to empirical evaluation with static requests, we test the system performance in a more realistic, large-scale scenario. We built a simulator in Python based on the results we obtained in the previous section. Since no suitable traces could be obtained to benchmark serverless workloads, our simulation used the Pareto distribution to emulate the inter-arrival time of request events. Wilson\cite{nwtraffic} suggests Pareto distribution to be a simple approximation of network events when the shape of the distribution is close to 1. We simulated 5 concurrent services, each having 1000 requests in total with inter-arrival times of a Pareto distribution with shape $= 1.1$. We assume that each service is independent, and a service only needs one instance to handle all the incoming traffic. 100 trials were run for both the short application (initialization 7s, scale down cooldown 30s) and long application (initialization 32s, scale down cooldown 60s) setups. Sample traces and corresponding CDFs are shown in figures \ref{fig:7_30_sample} and \ref{fig:32_60_sample}. The P95, P99 and P99.5 response time are shown in figures \ref{fig:7_30_percentiles} and \ref{fig:32_60_percentiles}.

We can see that the adding a pool significantly improves the overall response time. Even with 1 pod in the pool, the P95 for both short and long applications are entirely eliminated, and at least 50\% reduction for P99. Longer application initialization time increases the extreme percentiles. This is due to more requests pending when the instance takes longer to bootstrap.

\subsection{Response Time for Fixed Pool Size}
\begin{figure}[ht]
    \centering
    \includegraphics[width=0.48\textwidth]{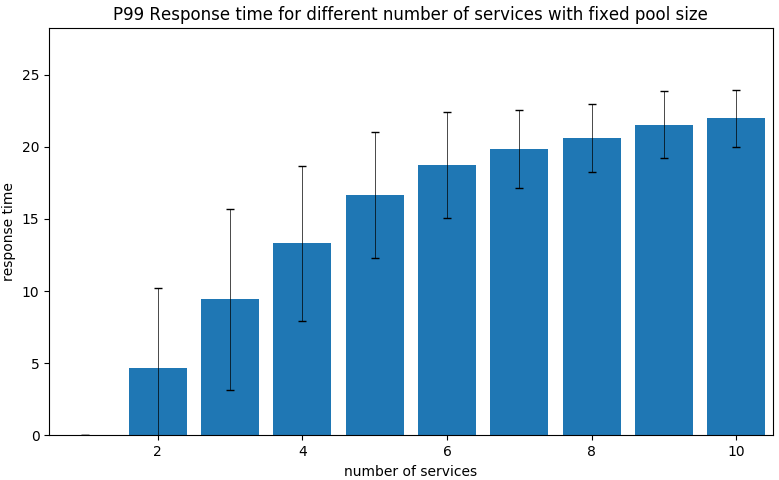}
    \caption{Response time with fixed pool size = 1}
    \label{fig:nsvc}
\end{figure}

In the experiments described in the previous section, we observe that even with one pod in the pool, the improvement is significant. We performed a simulation ranging between 1 to 10 services, while fixing the number of pods in the pool to 1 to see how contention affects the response time. The results are shown in Figure \ref{fig:nsvc}.

We can see that the improvement in P99 response time falls under 50\% if there are more than 5 services. However, the results of 1 to 5 services suggests that we would not need a large pool to reduce the cold start problem. This also shows that we do not need to bring in excessive resources for the pool when leveraging statistical multiplexing.

\section{Related Work}
The advent of serverless platforms has been less than five years, and since most details of serverless platforms originated from the industry, there is limited research in the realm. A prewarming mechanism in OpenWhisk is described in \cite{Thommes17}. However, the prewarming step only allocates the container runtime (e.g., a container with NodeJS), but not the function itself - which fits well stateless workloads, where functions do not have long  application initialization overhead. McGrath et al. \cite{McGrath17} also proposed a similar solution in Microsoft Azure with global cold queues that have pre-allocated empty containers ready to be initialized as any function, and a warm stack per function that reuses previous containers. This approach provides the flexibility to reduce platform-wise cold start overhead, but the application overhead cannot be mitigated. A recent work proposed by Akkus et al. \cite{Akkus18} turns to alternative methods instead of containers. They use forked processes as the function instance, allowing workloads to run with lower memory footprint and faster initialization time. However, not all programming languages support native forking, limiting the usage to particular languages such as Java and C/C++. Inter-function security is also a concern for process-based FaaSes due to lower degree of isolation between processes.

The cold start program can be also seen as a variation of known computer science models, as follows. 

Caching \cite{Vanderwiel00} is an active and related topic that has been explored for decades. Cache misses are costly when they happen, similar to the huge penalty in a cold start. The "prefetching/prewarming" concept also originates from cache designs. However, a fundamental difference between caching and our problem is that caching has two dimensions to utilize: temporal locality and spatial locality, while our problem only considers temporal locality. Therefore, many approaches leveraging spatial locality are invalid in our context. Caching also assumes that the benefits are shared among multiple users, i.e. entries in the cache can be accessed by all users requesting the same resource. In the cold start problem, however, instances are mutually exclusive and cannot be shared (for security purposes).

The knapsack problem appears to have similarities to cold starts, where we can model different functions with different values and optimizing the total value with limited resources. However, this model is a static approach not considering different time slots and the traffic distribution. The diminishing marginal return of value of multiple warm functions in the pool also cannot be accurately represented with static values.

Inventory models \cite{PORTEUS1990605} are stochastic models that model the behavior of inventory warehouses, widely used in operation research. By modeling each function as a type of "product" our system provides, our problem is essentially managing the inventory in a way that minimizes the overall time the clients wait starting from requesting a product until finish using the product. In our case, our model is similar to a multi-reparable-product single-facility model with holding costs\cite{PORTEUS1990605}. We can extend this formulation based on the system we are trying to mode, e.g. a multi-warehouse problem where we want to model deployments on a multi-server cluster.

\section{Future Work}
In a subsequent study on this problem, the following topics can be considered:
\begin{itemize}
    \item Sharing pools among multiple services should further improve cluster efficiency, while maintaining the latency improvement benefits.
    \item Simulation: Our simulation only covers scaling between 0 and 1. Scaling up multiple instances at a time could be interesting since contention between services would be even more significant. Another improvement would be simulating hybrid traces, i.e. traces in both short and long applications, and observe if the behaviour of different services interact with each other. The impact on resource utilization of the host system could also be taken into account when running the simulations.
    \item Theory: As mentioned in the previous section, mathematical models could be useful to obtain theoretical insights of the cold start problem. 
\end{itemize}
\section{Conclusion}
Mitigating the cold start problem is crucial for future serverless platforms. With a basic implementation and initial simulations, our solution using a pool of warm function instances suggests that the P99 cold start delay can be reduced by up to 85\% with one pod in the pool. Further research and experiments should be conducted in order to analyze the overall trade-offs of our approach.

\bibliographystyle{ACM-Reference-Format}

\begin{thebibliography}{00}


\ifx \showCODEN    \undefined \def \showCODEN     #1{\unskip}     \fi
\ifx \showDOI      \undefined \def \showDOI       #1{#1}\fi
\ifx \showISBNx    \undefined \def \showISBNx     #1{\unskip}     \fi
\ifx \showISBNxiii \undefined \def \showISBNxiii  #1{\unskip}     \fi
\ifx \showISSN     \undefined \def \showISSN      #1{\unskip}     \fi
\ifx \showLCCN     \undefined \def \showLCCN      #1{\unskip}     \fi
\ifx \shownote     \undefined \def \shownote      #1{#1}          \fi
\ifx \showarticletitle \undefined \def \showarticletitle #1{#1}   \fi
\ifx \showURL      \undefined \def \showURL       {\relax}        \fi
\providecommand\bibfield[2]{#2}
\providecommand\bibinfo[2]{#2}
\providecommand\natexlab[1]{#1}
\providecommand\showeprint[2][]{arXiv:#2}

\bibitem[\protect\citeauthoryear{??}{aws}{2014}]%
        {awslambda}
 \bibinfo{year}{2014}\natexlab{}.
\newblock \bibinfo{title}{AWS Lambda}.
\newblock   (\bibinfo{date}{Nov} \bibinfo{year}{2014}).
\newblock
\showURL{%
\url{https://aws.amazon.com/lambda/}}


\bibitem[\protect\citeauthoryear{??}{ope}{2016}]%
        {openwhisk}
 \bibinfo{year}{2016}\natexlab{}.
\newblock \bibinfo{title}{OpenWhisk}.
\newblock   (\bibinfo{year}{2016}).
\newblock
\showURL{%
\url{https://openwhisk.apache.org}}


\bibitem[\protect\citeauthoryear{??}{fir}{2018}]%
        {firecracker}
 \bibinfo{year}{2018}\natexlab{}.
\newblock \bibinfo{title}{AWS Firecracker}.
\newblock   (\bibinfo{date}{Dec} \bibinfo{year}{2018}).
\newblock
\showURL{%
\url{https://firecracker-microvm.github.io}}


\bibitem[\protect\citeauthoryear{??}{gcf}{2018}]%
        {gcf}
 \bibinfo{year}{2018}\natexlab{}.
\newblock \bibinfo{title}{Google Cloud Functions}.
\newblock   (\bibinfo{date}{Jul} \bibinfo{year}{2018}).
\newblock
\showURL{%
\url{https://cloud.google.com/functions/}}


\bibitem[\protect\citeauthoryear{??}{kna}{2018a}]%
        {knative}
 \bibinfo{year}{2018}\natexlab{a}.
\newblock \bibinfo{title}{Knative}.
\newblock   (\bibinfo{year}{2018}).
\newblock
\showURL{%
\url{https://cloud.google.com/knative}}


\bibitem[\protect\citeauthoryear{??}{kna}{2018b}]%
        {knativeserving}
 \bibinfo{year}{2018}\natexlab{b}.
\newblock \bibinfo{title}{Knative Serving}.
\newblock   (\bibinfo{year}{2018}).
\newblock
\showURL{%
\url{https://github.com/knative/serving}}


\bibitem[\protect\citeauthoryear{Akkus, Chen, Rimac, Stein, Satzke, Beck,
  Aditya, and Hilt}{Akkus et~al\mbox{.}}{2018}]%
        {Akkus18}
\bibfield{author}{\bibinfo{person}{Istemi~Ekin Akkus},
  \bibinfo{person}{Ruichuan Chen}, \bibinfo{person}{Ivica Rimac},
  \bibinfo{person}{Manuel Stein}, \bibinfo{person}{Klaus Satzke},
  \bibinfo{person}{Andre Beck}, \bibinfo{person}{Paarijaat Aditya}, {and}
  \bibinfo{person}{Volker Hilt}.} \bibinfo{year}{2018}\natexlab{}.
\newblock \showarticletitle{{SAND}: Towards High-Performance Serverless
  Computing}. In \bibinfo{booktitle}{{\em 2018 {USENIX} Annual Technical
  Conference ({USENIX} {ATC} 18)}}. \bibinfo{publisher}{{USENIX} Association},
  \bibinfo{address}{Boston, MA}, \bibinfo{pages}{923--935}.
\newblock
\showISBNx{978-1-931971-44-7}
\showURL{%
\url{https://www.usenix.org/conference/atc18/presentation/akkus}}


\bibitem[\protect\citeauthoryear{Childers}{Childers}{2018}]%
        {childers18}
\bibfield{author}{\bibinfo{person}{Chip Childers}.}
  \bibinfo{year}{2018}\natexlab{}.
\newblock \bibinfo{title}{Why Companies Are Adopting Serverless Cloud
  Technology}.
\newblock   (\bibinfo{date}{May} \bibinfo{year}{2018}).
\newblock
\showURL{%
Retrieved Dec 12, 2018 from
  \url{https://www.forbes.com/sites/forbestechcouncil/2018/05/18/why-companies-are-adopting-serverless-cloud-technology}}


\bibitem[\protect\citeauthoryear{McGrath and Brenner}{McGrath and
  Brenner}{2017}]%
        {McGrath17}
\bibfield{author}{\bibinfo{person}{G. McGrath} {and} \bibinfo{person}{P.~R.
  Brenner}.} \bibinfo{year}{2017}\natexlab{}.
\newblock \showarticletitle{Serverless Computing: Design, Implementation, and
  Performance}. In \bibinfo{booktitle}{{\em 2017 IEEE 37th International
  Conference on Distributed Computing Systems Workshops (ICDCSW)}}.
  \bibinfo{pages}{405--410}.
\newblock
\showISSN{2332-5666}
\showDOI{%
\url{https://doi.org/10.1109/ICDCSW.2017.36}}


\bibitem[\protect\citeauthoryear{Porteus}{Porteus}{1990}]%
        {PORTEUS1990605}
\bibfield{author}{\bibinfo{person}{Evan~L. Porteus}.}
  \bibinfo{year}{1990}\natexlab{}.
\newblock \showarticletitle{Chapter 12 Stochastic inventory theory}.
\newblock In \bibinfo{booktitle}{{\em Stochastic Models}}.
  \bibinfo{series}{Handbooks in Operations Research and Management Science},
  Vol.~\bibinfo{volume}{2}. \bibinfo{publisher}{Elsevier}, \bibinfo{pages}{605
  -- 652}.
\newblock
\showISSN{0927-0507}
\showDOI{%
\url{https://doi.org/10.1016/S0927-0507(05)80176-8}}


\bibitem[\protect\citeauthoryear{Th$\ddot{o}$mmes}{Th$\ddot{o}$mmes}{2017}]%
        {Thommes17}
\bibfield{author}{\bibinfo{person}{Markus Th$\ddot{o}$mmes}.}
  \bibinfo{year}{2017}\natexlab{}.
\newblock \bibinfo{title}{Squeezing the milliseconds: How to make serverless
  platforms blazing fast!}
\newblock   (\bibinfo{date}{Apr} \bibinfo{year}{2017}).
\newblock
\showURL{%
Retrieved Dec 13, 2018 from
  \url{https://medium.com/openwhisk/squeezing-the-milliseconds-how-to-make-serverless-platforms-blazing-fast-aea0e9951bd0}}


\bibitem[\protect\citeauthoryear{Tresness}{Tresness}{2018}]%
        {tresness18}
\bibfield{author}{\bibinfo{person}{Colby Tresness}.}
  \bibinfo{year}{2018}\natexlab{}.
\newblock \bibinfo{title}{Why Companies Are Adopting Serverless Cloud
  Technology}.
\newblock   (\bibinfo{date}{Feb} \bibinfo{year}{2018}).
\newblock
\showURL{%
Retrieved Dec 12, 2018 from
  \url{https://blogs.msdn.microsoft.com/appserviceteam/2018/02/07/understanding-serverless-cold-start}}


\bibitem[\protect\citeauthoryear{Vanderwiel and Lilja}{Vanderwiel and
  Lilja}{2000}]%
        {Vanderwiel00}
\bibfield{author}{\bibinfo{person}{Steven~P. Vanderwiel} {and}
  \bibinfo{person}{David~J. Lilja}.} \bibinfo{year}{2000}\natexlab{}.
\newblock \showarticletitle{Data Prefetch Mechanisms}.
\newblock \bibinfo{journal}{{\em ACM Comput. Surv.\/}} \bibinfo{volume}{32},
  \bibinfo{number}{2} (\bibinfo{date}{June} \bibinfo{year}{2000}),
  \bibinfo{pages}{174--199}.
\newblock
\showISSN{0360-0300}
\showDOI{%
\url{https://doi.org/10.1145/358923.358939}}


\bibitem[\protect\citeauthoryear{Wilson}{Wilson}{[n. d.]}]%
        {nwtraffic}
\bibfield{author}{\bibinfo{person}{Michael Wilson}.} \bibinfo{year}{[n.
  d.]}\natexlab{}.
\newblock \bibinfo{title}{A Historical View of Network Traffic Models}.
\newblock   (\bibinfo{year}{[n. d.]}).
\newblock
\showURL{%
\url{https://www.cse.wustl.edu/~jain/cse567-06/ftp/traffic_models2}}


\end{thebibliography}

\end{document}